\documentclass[12pt,letterpaper]{article}
\usepackage[a4paper, total={7in, 10in}]{geometry}

\usepackage{graphicx}
\usepackage{helvet}
\usepackage{authblk}
\usepackage{hyperref}
\usepackage{amsmath} 
\usepackage{amssymb} 
\usepackage{orcidlink} 
\usepackage[super,comma,sort&compress]  
   {natbib}

\usepackage{subcaption}
\usepackage{makecell}
\usepackage{booktabs}
\usepackage{nameref}  
\makeatletter
\renewcommand{\maketitle}{\bgroup\setlength{\parindent}{0pt}
\begin{flushleft}
  \textbf{\@title}
  
  \@author
\end{flushleft}\egroup}
\makeatother

\title{Frequency-Specific Neural Response and Cross-Correlation Analysis of Envelope Following Responses to Native Speech and Music Using Multichannel EEG Signals: A Case Study}
\date{}

\author[1,5,* \orcidlink{0000-0003-2612-7248}]{Md. Mahbub Hasan}
\author[2,3 \orcidlink{0000-0003-2565-5321}]{Md Rakibul Hasan}
\author[2 \orcidlink{0000-0003-1892-831X}]{Md Zakir Hossain}
\author[2,4 \orcidlink{0000-0001-8356-4909}]{Tom Gedeon}

\affil[1]{Department of Electrical and Electronic Engineering, Khulna University of Engineering \& Technology, Khulna 9203, Bangladesh}
\affil[2]{School of Electrical Engineering, Computing and Mathematical Sciences, Curtin University, Bentley WA 6102, Australia}
\affil[3]{Department of Electrical and Electronic Engineering, BRAC University, Dhaka 1212, Bangladesh}
\affil[4]{Obuda University, Budapest, Hungary.}

\affil[5]{Lead contact}

\affil[*]{Correspondence: mahbub01@eee.kuet.ac.bd}

\begin{document}

\maketitle

\section*{Summary}

Although native speech and music envelope following responses (EFRs) play a crucial role in auditory processing and cognition, their frequency profile, such as the dominating frequency and spectral coherence, is largely unknown. We have assumed that the auditory pathway -- which transmits envelope components of speech and music to the scalp through time-varying neurophysiological processes -- is a linear time-varying system, with the envelope and the multi-channel EEG responses as excitation and response, respectively. This paper investigates the transfer function of this system through two analytical techniques -- time-averaged spectral responses and cross-spectral density -- in the frequency domain at four different positions of the human scalp. Our findings suggest that \emph{alpha (8-11 Hz)}, \emph{lower gamma (53-56 Hz)}, and \emph{higher gamma (78-81 Hz)} bands are the peak responses of the system. These frequently appearing dominant frequency responses may be the key components of familiar speech perception, maintaining attention, binding acoustic features, and memory processing. The cross-spectral density, which reflects the spatial neural coherence of the human brain, shows that \emph{10-13 Hz}, \emph{27-29 Hz}, and \emph{62-64 Hz} are common for all channel pairs. As neural coherences are frequently observed in these frequencies among native participants, we suggest that these distributed neural processes are also dominant in native speech and music perception. 

\section*{Keywords}


Envelope following response, Time-varying system, Transfer function, Dominant spectral response, Cross-spectral density, EEG

\section*{Introduction}
Speech and Music envelope following responses (EFRs) in the human brain are crucial to understanding auditory processing and cognition, and correspond to the brain's ability to track the slow amplitude modulations of speech signals. Electroencephalography (EEG), a non-invasive neural potential recording technique, is commonly used to capture EFRs and estimate the hearing threshold. EFRs are frequency-following responses of the spectral components of speech and music that propagate through the auditory system, beginning at the ear and progressing through the brainstem (e.g.,  cochlea, hair cells, superior olivary complex, lateral lemniscus, inferior colliculus), and medial geniculate body of the thalamus, before reaching the primary auditory cortex. From there, these components reach the scalp via volume conduction. The entire propagation path is called the auditory pathway. The envelope components propagate by mechanical, electro-mechanical, and electro-chemical processes, which are time-varying in nature. These time-varying neurophysiological processes within the auditory pathway can be modelled as a linear system. Crosse et al. reviewed the aspects of representing the neurophysiological process by a linear time-varying system \cite{crosse2021linear}. The path offers time and frequency-dependent gain and phase-shifting angle to the envelope spectral components. For the time-dependent nature of this auditory pathway, the temporal transfer function exhibits the dynamic characteristics of the pathway.

A transfer- or response-function is a mathematical representation that characterises the excitation-output relationship of the auditory pathway. In the context of EFRs, it relates the envelope (excitation) to the corresponding neural response (output) captured by EEG. Understanding the transfer- or response-function of the EEG channels in response to speech envelopes provides a crucial window into the neural dynamics of speech perception and processing. The transfer- or response-function exhibits how different brain regions respond to the temporal features of speech/ music and reveals underlying neural mechanisms \cite{teng2021modulation}. This analysis has clinical significance, as it can identify EEG biomarkers for diagnosing and treating conditions such as hearing impairment, auditory processing disorders, and language disorders \citep{chandrasekaran2009context}. Furthermore, estimating the transfer- or response-function facilitates the assessment of cortical connectivity patterns during speech processing, thereby deepening our understanding of how different brain regions interact to process auditory information \citep{boucsein2011beyond}. The transfer function reflects the coupling between the stimulus envelope and its response can reveal the underlying processes of speech-in-noise perception \cite{mchaney2021cortical}, and selective auditory attention \cite{baltzell2016attention}. Numerical investigation of the transfer function contributes to the refinement of the computational models of neural representation of speech intelligence \cite{macintyre2024neural}, and has implications in cochlear implant design \cite{nogueira2022predicting}. Scalp-recorded EEG potentials in the low-frequency range (approximately 0-100 Hz) reflect the neural responses to the auditory stimulus envelope \cite{giraud2012cortical}. Stimuli with a duration of more than 500 ms generally elicit steady-state or sustained responses. The time-average transfer function of such time-varying channels exhibits the key dynamic properties of the auditory pathway \cite{picton2003human,david2022evaluation}.

The syllabic and phonemic spectrum is associated with the time-averaged frequency-dependent spectrum within 0-100 Hz. \citet{picton2003human} reported 40 Hz as the dominant auditory steady state response (aSSR) for waking subjects. Although the EFR to sinusoidally amplitude-modulated stimulus decreases with frequency, significant responses were exceptionally observed around 40 Hz and 80-100 Hz \cite{gransier2021stimulus}. The spectral responses that sustained throughout the presentation of long auditory excitation are the most dominant in the time-averaged responses. Although the aSSR for amplitude-modulated pure tone provides frequency-specific information, the response is weak. However, the aSSR can be strengthened by expanding the bandwidth, which suggests the use of natural speech or music as stimuli \citep{sergeeva2024effect}.
By applying the inverse Fourier transform on the time-averaged frequency response, \citet{kulasingham2024predictors} have estimated the temporal response function (TRF)--a time-domain representation of the auditory pathway's temporal impulse response. Unlike the frequency domain transfer- or response-function, the TRF explicitly describes the time-dependent dynamics of the auditory pathway. The TRF is computationally analogous to the auditory brainstem response, but it is estimated using steady-state stimuli. An alternative method to using the inverse Fourier transform is the multivariate TRF analysis, which extends linear regression to relate continuous stimuli to neural responses \citep{crosse2016the}. Such estimation of transfer functions using linear models assumes that the convolution of the temporal response function with excitation produces the system's response/output. Additionally, the temporal modulation transfer function (TMTF) is utilised to represent the response of the auditory pathway, which is the transfer function estimated by a single-frequency amplitude-modulated signal used as an excitation \cite{parida2024rapid}. The TMTF reflects the response related to discrete specific modulation frequencies. It reveals the minimum detectable amplitude fluctuation of the participant, which also indicates the hearing threshold \cite{dimitrijevic2016human}.

Cross-correlation and coherence analyses assess the interaction and connectivity among brain regions engaged in speech perception \citep{bidelman2015multichannel}. High cross-correlation at zero lag between channels indicates a strong, direct connection between the corresponding brain regions, which helps in mapping functional networks involved in speech processing \citep{nunez2006electric}. The frequency-specific neural coherence among different brain regions is investigated for epilepsy patients stimulated by music and speech by \citet{te2024speech}. The spatial neural coherence between the left and right frontal areas of the human brain increases in theta, alpha and gamma frequency bands with musical learning, as reported by \citet{peterson2007music}. The EEG responses align with the temporal dynamics of speech signals and thus enhance our understanding of how the brain tracks and processes speech \citep{ding2012emergence}. Like transfer function analysis, these methods have clinical relevance for studying speech perception deficits in populations with auditory processing disorders or language impairments \citep{abrams2008right}.

Musical familiarity enhances the neural synchronisation at a lower frequency level (1-4 Hz). The participants score their familiarity numerically, ranging from $-100$ to $+100$. The relation between familiarity and the temporal response function is investigated by \citet{Weineck2022}. \citet{DiLiberto2021Neural} investigated the relationship between language proficiency and temporal response function. Our previous work \cite{zhou2024how} investigated the dominant frequency components, phase locking value, and weighted phase locked index for native and non-native speakers. The relationship between neural responses and auditory stimuli, described by the transfer function of the auditory neural pathway, remains under investigation and is not yet well documented. To fully leverage the insights gained from the transfer function for native speech/music excitation, in this article, we adopt two analytical techniques: time-averaged spectral responses and cross-spectral density or spectral coherence among the different regions of the human scalp.

\section*{Results and discussion}
\subsection*{Stimuli and Responses}
\begin{figure}[!htb]
  \centering
  \begin{subfigure}{0.49\textwidth}
    \centering
    \includegraphics[width=\linewidth]{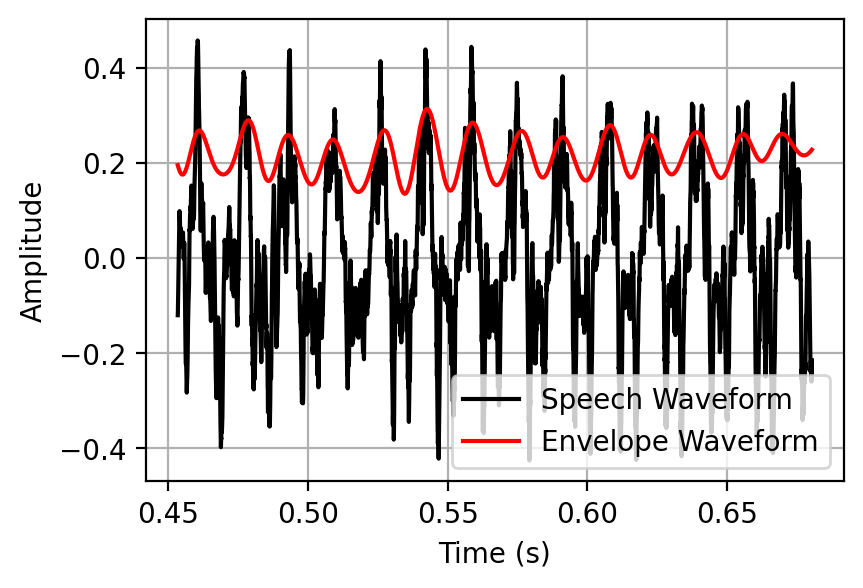}
    \caption{Time domain}
    \label{subfig:speech_env_t}
  \end{subfigure}\hfill
  \begin{subfigure}{0.49\textwidth}
    \centering
    \includegraphics[width=\linewidth]{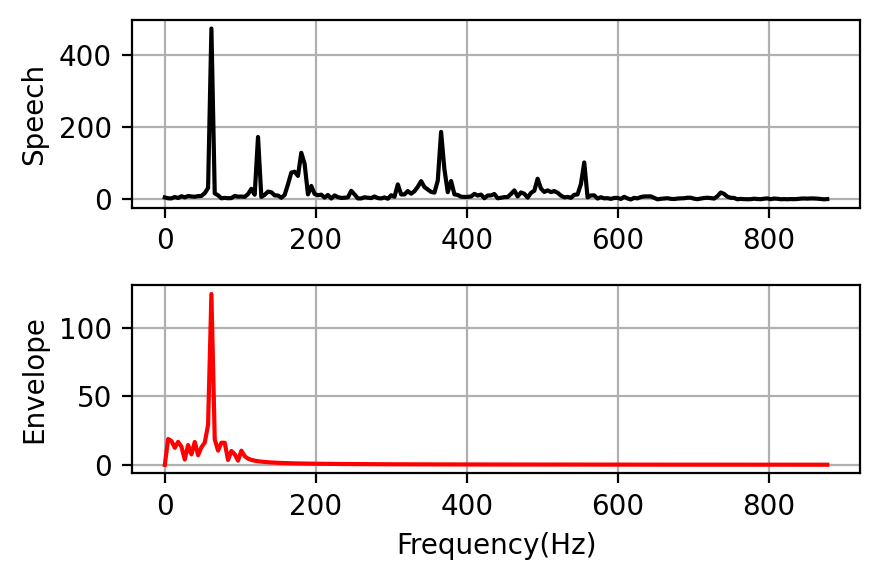}
    \caption{Frequency domain (top: speech waveform; bottom: envelope waveform)}
    \label{subfig:speech_env_f}
  \end{subfigure}
  \caption{Estimation of envelope function from stimulus signal. The envelope function is the absolute value of the low-pass filtered analytic signal of stimulus. The speech envelope function focuses on slower modulations or fluctuations in the stimulus signal. Here, the range is 0--100 Hz.}
  \label{fig:speech_env}
\end{figure}

An example of a stimulus waveform (A detailed description of stimuli is provided later in Section \nameref{sec:database}) and its corresponding envelope is presented in \autoref{fig:speech_env}. \autoref{subfig:speech_env_t} illustrates time-domain signals, where the stimulus waveform shows rapid fluctuations -- typical of a speech signal. The envelope waveform represents a smoothed version of the original speech waveform. This envelope preserves the amplitude variations over time, thus emphasising the modulation patterns within the speech signal. The envelope waveform features periodicity and relates to the underlying phonetic or syllabic structure. \autoref{subfig:speech_env_f} presents the speech waveform's spectrum (top) and the envelope's spectrum (bottom). The speech waveform spectrum shows prominent peaks at various frequencies. The most significant peak is located at a low frequency, followed by additional peaks at higher frequencies. The envelope waveform spectrum exhibits a significant peak around the same low frequency as seen in the speech spectrum. The higher frequencies, however, show significantly reduced amplitude, which indicates that the envelope predominantly captures the lower frequency modulations of the original stimulus signal.

\begin{figure}[t!]
    \centering
    \includegraphics[width=0.7\linewidth]{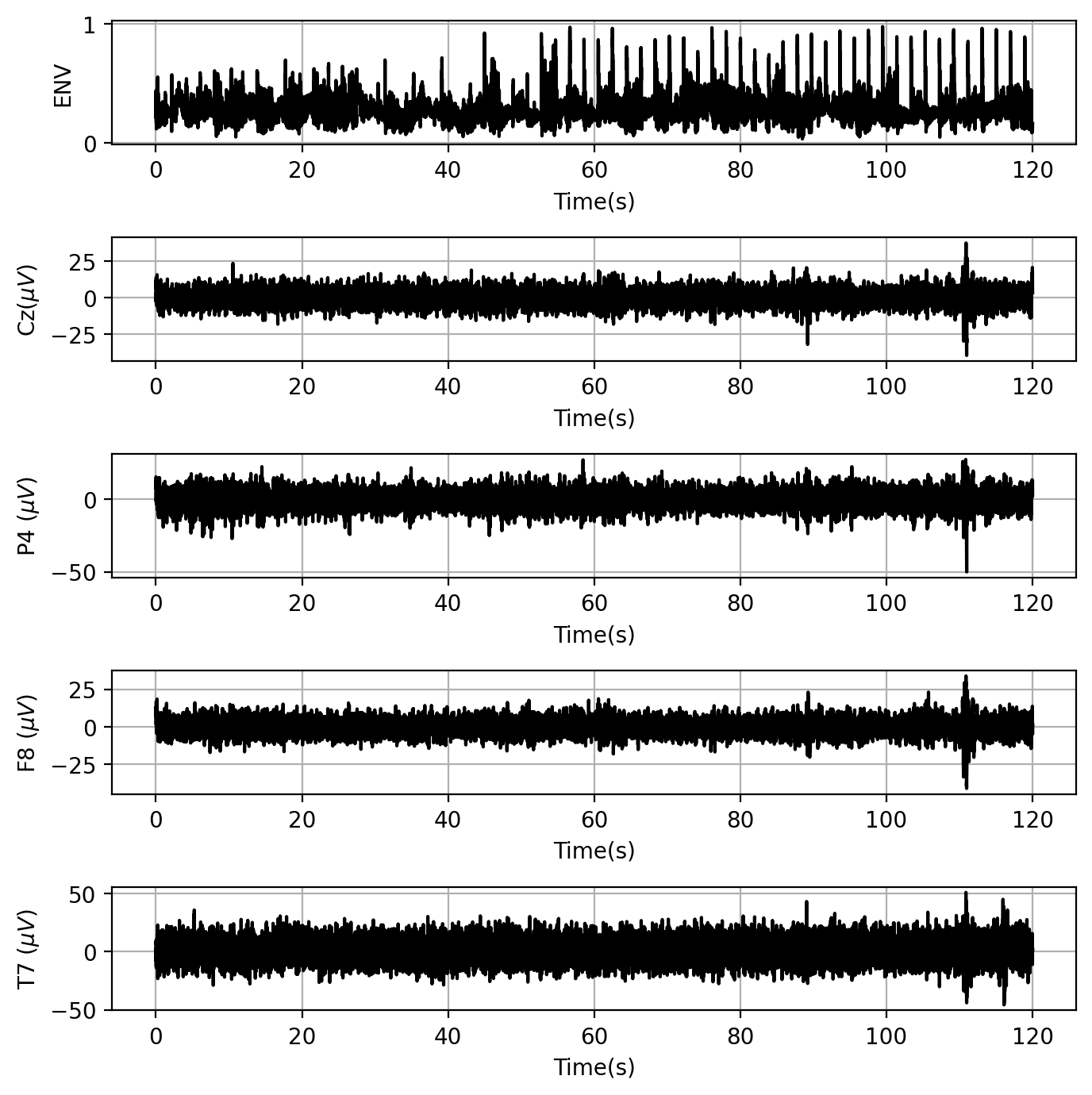}
    \caption{Auditory evoked potential captured at different electrodes with envelope (topmost).}
    \label{fig:ENV_EEGs}
\end{figure}

The speech- and music-stimulated EEG signals (A detailed description of EEG signal is given later in Section \nameref{sec:database}) were recorded from electrodes placed at four different locations on the scalp: Cz (central), P4 (right parietal), F8 (right frontal) and T7 (left temporal). \autoref{fig:ENV_EEGs} shows these signals with speech envelope, which reflect the brain’s electrical activity in response to the speech stimulus. The EEG response signal contains frequency components up to 100 Hz, consistent with the Nyquist criterion for a 200 Hz sampling rate, which states that the maximum representable frequency must be less than half the sampling frequency. According to linear systems theory and the frequency-following response (FFR) model, the 0–100 Hz components of the response EEG signal are assumed to be phase-locked to the corresponding 0–100 Hz components of the stimulus envelope, estimated from the composite speech and music signal. Based on this rationale, the recorded EEG signals are considered as EFRs.

The potential at Cz lies within $\pm 25 \mu V$ for the majority of the time; however, it occasionally shows brief deviations beyond the range. The Cz electrode, positioned at the vertex of the central scalp, records electrical activity from central brain regions, including the primary auditory cortex, which is involved in the processing of auditory stimuli. The neural activity is related to processing auditory features like pitch, rhythm, and amplitude modulations \cite{giraud2012cortical}. Like Cz, the potential at P4 has a similar range, with short deviations that include higher-magnitude negative peaks. This signal reflects the electrical activity generated by neuronal processes within the right parietal cortex. This region is involved with a range of cognitive functions, including spatial attention, sensory integration and the processing of abstract auditory information. Moreover, the neuronal processes of this region may include the integration of auditory stimuli with other sensory modalities and attentional processes. \cite{macaluso2005multisensory}. At F8, the EEG signal consists of neural activity from the right inferior frontal gyrus, particularly areas implicated in auditory functions, and has a similar amplitude range like Cz. The frontal regions process continuous auditory stimuli and are involved in attention and working memory \cite{friederici2011brain}. The left temporal EEG electrode, T7, captures neural activity from the superior temporal gyrus and surrounding regions. The amplitude of the potential at T7 is similar to that at P4. This left temporal region processes the phonological aspects and comprehension \cite{hickok2007cortical, friederici2011brain}.

 \begin{figure}[t!]
    \centering
    \includegraphics[width=0.7\linewidth]{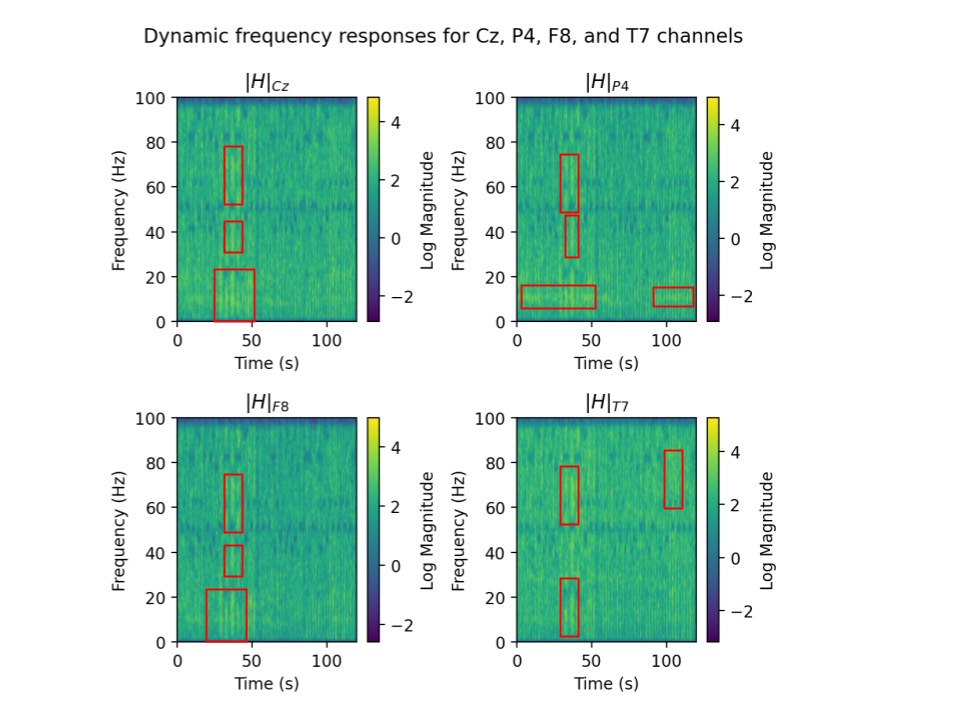}
    \caption{Time-frequency representations of the frequency responses for Cz, P4, F8 and T7 Channels. Each subplot shows an intensity plot of the log magnitude of the frequency response for the respective channel.}
    \label{fig:Dynamic_response}
\end{figure}
\subsection*{Temporal frequency response of the auditory pathways}
As the evoked potential generation system is time-varying, the frequency response of the evoked EEGs and envelope function has been estimated utilising Short-time Fourier transform (STFT) with a window length of 1 s and a hop length of 0.5 s (time resolution). The dynamic frequency spectrum $(log_{10}(|H(m,k)|)$ (estimated using \autoref{eqn:TF}, to be shown later) of channels Cz, P4, F8 and T7 for speaker (s04) are displayed in \autoref{fig:Dynamic_response} using time-frequency intensity plots. It shows the changes over the 0--100 Hz frequency range and across the 0--120 s time range. The cortical and sub-cortical tracking of the dynamic features (modulation rate, frequency contents, etc.) of the stimuli yields the fluctuating frequency responses \cite{Aiken2008, ZionGolumbic2013}. 

For the Cz channel, the dominating 0--20 Hz frequency response appears within 25--50 seconds. Within the 0--20 Hz range, the delta (0--4 Hz), theta (4--8 Hz), and low beta (13--20 Hz) frequency bands exhibit dominant responses, consistent with the findings of the MEG-based study \cite{luo2007phase}. Higher frequency dominating bands emerged at around 35--42 Hz and 53--78 Hz within 30--40 seconds. For the P4 channel, the 4--14 Hz frequency bands dominate at the 2--50 s and 90--120 s time intervals. Two additional dominating frequency bands, 30--46 Hz and 48--70 Hz, become visible within the approximate 30--40 s interval. The parietal P4 channel exhibits prominent responses in the theta (4--8 Hz) and alpha (8--12 Hz) bands, with amplitude variations over time. These frequency bands are associated with attentional engagement and cognitive efforts \cite{klimesch1999eeg}. The variations of frequency response in the theta and alpha over time indicate likely dynamic shifts in attentional engagement and cognitive control \cite{mcgarrigle2014listening}. The F8 channel shows a dominating frequency similar to that of the Cz, but its amplitude is lower than that of the Cz. The frequency response of the right frontal, F8, contains a broader spectrum across the beta (13--30 Hz) and gamma (30--50 Hz and 60--80 Hz) bands. The beta and gamma band oscillations are varied according to auditory attention tasks in the frontal cortex. The increased power in these bands indicates greater attentional focus at the time of distinguishing between relevant and irrelevant auditory features of stimuli \cite{giraud2012cortical}. These time-dependent variations of frequency response may be associated with the continuous adjustment of neural activities to maintain focus on stimuli, especially segregating speech from noise. For the T7 channel, the two dominating frequency bands, 2-22 Hz and 52-72 Hz, manifested within approximately 30–40 seconds. Another dominating frequency band, 60-80 Hz, arose within approximately 100-110 seconds. These frequency bands in the left temporal region follow the rhythmic properties of auditory stimulus dynamically \cite{giraud2012cortical}. The maximum dominating frequency components for all four channels lie within 20--50 s. The channels P4 and T7 exhibit additional dominating frequency responses near 100 s. The time duration of dominating frequency responses varies across all channels due to the propagation and dynamics of each channel path. The dominating frequencies of Cz and F8 channels are similar. The range of the dominating frequencies is approximately range from 0--80 Hz. However, within this frequency range, Cz and F8 exhibit broader frequency responses than the other two channels. The frequency tuning of different brain regions or channels may cause the varying dominating frequencies across channels. The transfer functions of the four EEG channels, stimulated by the speech envelope, exhibit time-varying characteristics. These modulations occur in response to neural activities, enabling the brain to track and encode the temporal and spectral features of the speech envelope. The reasons for being time-varying are the multisensory integration \cite{xi2020characterizing} of the human brain and noise level \cite{borges2024speech}, both of which change over time. In this article, we aim to estimate the dominating frequency components and the frequency components that exhibit strong inter-channel cross-correlation in a 120-second stimulus. These estimates correspond to the steady-state transfer function or frequency response. We averaged the temporal transfer function to determine the dominant frequency band or range of the channel transfer function to obtain a frequency response. The averaging may reduce the interference of other sensors and noise.

\subsection*{Time-averaged multi-channel frequency and correlation }
\begin{figure}[t!]
    \centering
    \includegraphics[width=0.8\linewidth]{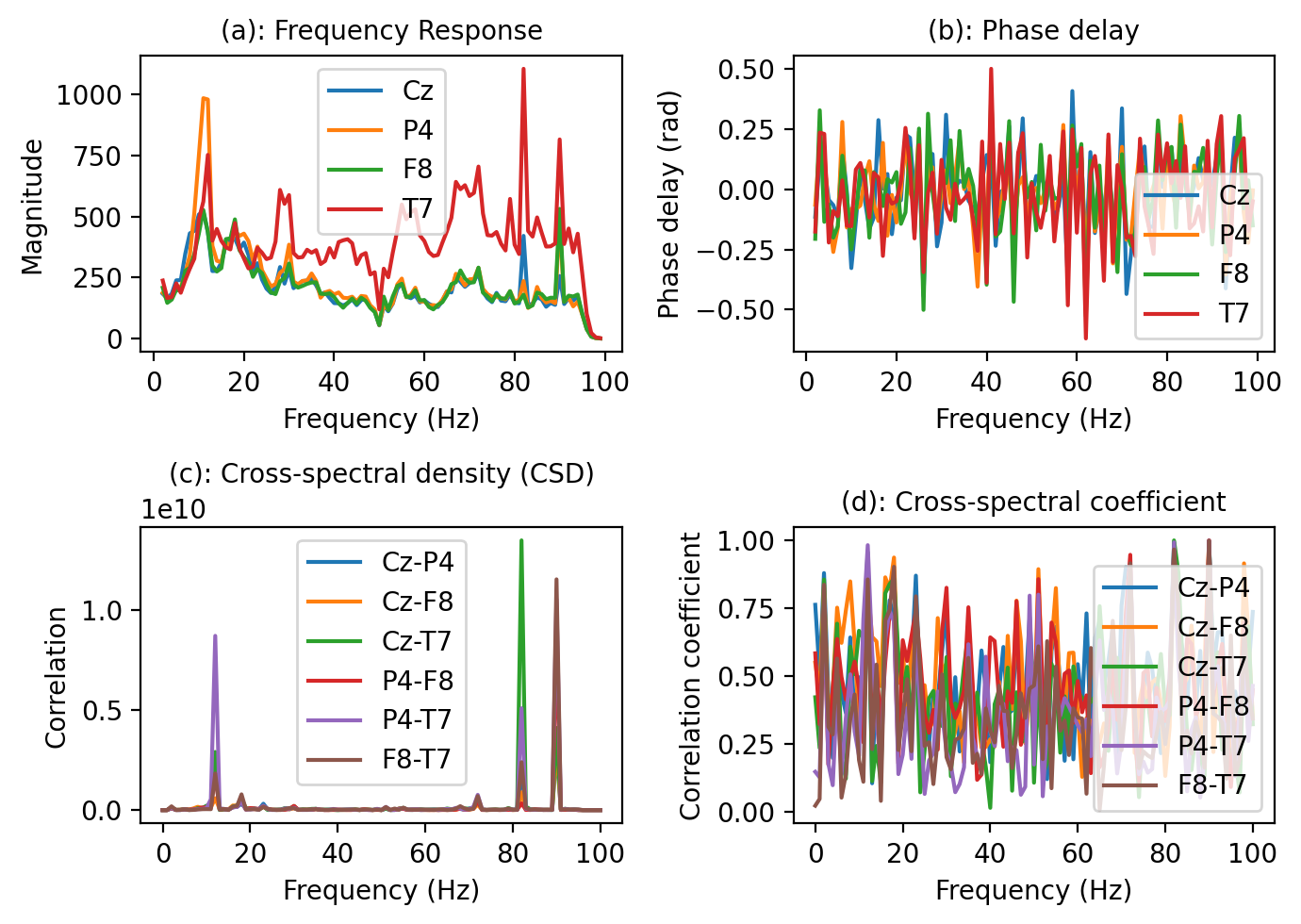}
    \caption{Frequency-dependent speech envelope following responses (EFRs) across four EEG channels: Cz, P4, F8 and T7. (a) The time-average frequency response magnitude determines how different frequency components contribute to the EEG signal at different channels, with distinct peaks indicating dominant frequency bands. (b) The phase delay provides insight into the relative timing and coordination of auditory evoked potentials of these channels. (c) Cross-spectral density between channel pairs measures the degree of synchronisation or coupling between the channels, with peaks suggesting strong inter-channel coordination at specific frequencies. (d) Cross-spectral coefficients indicate the coupling between the channels across the frequencies without considering the magnitude of the frequency components.}
    \label{fig:comb}
\end{figure}

For the estimation of the dominating frequency components and the inter-channel cross-correlation, we have calculated four metrics: time-averaged frequency response ($ \frac{1}{M} \sum_{m=0}^{M-1}|H(m,k)|$), phase delay ($\overline{\Delta \phi( k)}$), zero-lag CSD ($|R_{xy}(0,k)|$), and cross-correlation coefficients ($C_{xy}(k)$). The time-averaged frequency response provides a consistent profile of neural resonance frequencies related to auditory tracking \cite{luo2007phase}. The phase delay (phase differences between EFR and auditory envelopes) reflects the timing of neural tracking, adaptation, and predictive mechanisms \cite{gross2013speech}. Cross-spectral density (CSD) reflects the synchronous oscillatory energy shared between two EEG channels, highlighting potential functional connectivity between them in auditory perception \cite{gross2013speech}. The frequency-dependent cross-correlation coefficients are the normalised CSD, revealing the phase coupling between two time-varying channels without considering the magnitude of the frequency components. These four frequency-dependent metrics for an individual participant are shown in \autoref{fig:comb}.

\autoref{fig:comb} (a) illustrates the magnitude of the time-averaged frequency response across different channels (Cz, P4, F8 and T7) as a function of frequency. The response can reveal how the neural activities (recorded by EEG channels) of the brain are tuned to different frequency components of the envelope of the sound stimuli. The tuned frequency components having the peak magnitude in frequency response trajectories are related to the neural processing of specific frequency components of the envelope of auditory stimuli. Due to the interconnected nature of human brain regions, the dominant frequency responses of the transfer function for the EEG channels may exhibit similar characteristics when processing auditory stimuli \cite{hipp2012large}. It is known that each dominant frequency or frequency band may reflect distinct cognitive processes. Furthermore, when similar frequency responses appear across different brain regions, they may work in coordination to represent a perceptual or cognitive experience \cite{adamos2018harnessing}. The frequency response at Cz has several peaks: a prominent peak occurs at 11 Hz with a magnitude of 518.68 and a width of 99 Hz. We have identified two additional notable peaks at 18 Hz (magnitude: 409.33, width: 37 Hz) and 82 Hz (magnitude: 420.48, width: 49 Hz). At 28 Hz and 30 Hz, minor peaks are visible. The peak at 11 Hz at Cz is located within the alpha frequency band (8-12 Hz), strongly linked with sensory processing and attentional control. The focus on the auditory stimuli and background noise suppression processes modulates the alpha-band activity. The alpha-band activity plays a role in auditory perception, engaging the above neural processes \cite{foxe2011role}. In the P4 channel, several peaks are observed, such as a distinct peak at 11 Hz (within the alpha band) with a magnitude of 983.24 and a width of 99 Hz. Besides this, the response exhibits additional peaks in beta (at 18 Hz, magnitude: 440.92, width: 35 Hz) and gamma bands (at 90 Hz, magnitude: 485.33, width: 49 Hz). The beta response is associated with auditory prediction and temporal integration, aiding in the perception of speech rhythms and the temporal flow of auditory information \cite{chang2018beta}. Additionally, peaks at 23, 30 and 72 Hz, which fall within the beta and gamma frequency bands, indicate the involvement of higher frequencies in speech processing. The right frontal channel, F8, displays various frequency peaks. The main peak is observed at 90 Hz, exhibiting a magnitude of 530.84 and a width of 99 Hz. The right frontal cortex, known for its role in multisensory integration, may utilise the 90 Hz gamma frequency component to integrate auditory stimulus information with other sensory modalities \cite{senkowski2008crossmodal}. Among other significant peaks are 11 Hz (magnitude: 526.11, width: 50 Hz) and 18 Hz (magnitude: 487.90, width: 36 Hz). The T7 channel shows the highest number of significant peaks among the four channels. The most notable peak occurs at 82 Hz with a magnitude of 1103.40 and a width of 99 Hz. In the gamma frequency band, we identified an additional peak at 90 Hz, with a magnitude of 814.58 and a bandwidth of 12 Hz. We found another significant peak at 12 Hz (amplitude: 751.53, bandwidth: 50 Hz), indicating alpha band neural activity.

\autoref{fig:comb} (b) shows the phase delay across frequencies for each EEG channel. Phase delay, in this context, refers to the time shift between the stimulus (speech envelope) and the neural response recorded at each channel. A positive phase delay indicates that the neural response lags behind the stimulus, while a negative phase delay would suggest an anticipatory response \cite{Machado2020PhaseBistability}. The variability in phase delay across frequencies and channels suggests that different frequency components of the speech envelope are processed with different latencies. The propagation time of neural signals, the nature of the neural encoding of the stimulus \cite{Doelling2021ThetaBand}, and the neural adaptation and attention affect the phase delay \cite{Mohammadi2023PhaseLocking}. The randomness and fluctuations in phase delay may also highlight the processing of different frequency components by different time-varying neural circuits. 

The \autoref{fig:comb} (c) illustrates the cross-correlation between different EEG channels as a function of frequency. It shows the similarities of the frequency responses of the two EEG channels, considering the envelope function of the auditory stimulation. High cross-correlation values at specific frequencies reveal that the speech envelope's corresponding frequency components are similarly propagated through the two channels, suggesting some functional connectivity between the corresponding brain regions. The peaks in the cross-correlation plot might indicate specific frequencies at which corresponding brain areas are particularly synchronised or coupled in their processing of the speech envelope. 

\autoref{fig:comb} (d) represents the cross-correlation coefficients between different EEG channel pairs with frequency. The cross-correlation coefficient is a normalised measure of the coupling between different brain regions at specific frequencies. The fluctuations in cross-correlation coefficients across frequencies are not uniform across all frequencies, potentially reflecting the inconsistencies in propagation delays or phase shifts of these regions in processing different frequency components of the speech envelope. Since the cross-correlation coefficient depends only on the consistency of the phase delay between the transfer functions of the two channels, rather than on their amplitude, a frequency with lower responses in both transfer functions may still produce a higher correlation coefficient. Future research may focus on the cross-correlation, rather than its coefficient, to avoid the lower dominating frequency responses in brain regional synchronisation.

Overall, the \autoref{fig:comb} offers a detailed overview of the frequency response, phase delay, cross-correlation, and cross-correlation coefficients for a single participant. We estimated the most frequently dominant frequency responses in the transfer function and its cross-correlations by considering 13 participants in this study. Across participants, specific frequency ranges eliciting stronger neural responses appear in multiple participants. This consistency suggests that the neural processing mechanism underlying the perception of the speech envelope operates within these frequencies or frequency ranges. When scrutinising the cross-correlation between different EEG channels, we observed that certain frequency bands consistently exhibited high correlations across participants, suggesting similar functional connectivity patterns between brain regions.

\subsection*{Distribution of dominant frequency responses across participants}
Supplementary Table S1 lists the specific frequencies at which the stimulus-envelope response neural generator produces the maximum responses for each of the 13 participants. The information about the participants is given later in section \nameref{sec:database}. These peak frequencies were determined using the peak-fitting algorithm of SciPy Python package, with the identification criteria set at a threshold of 50\% of the maximum magnitude observed in the frequency response, and a minimum separation of 10 Hz between consecutive peaks. This estimation of distinct, significant peaks provides a clear indication of the dominant frequency responses for each participant. 

\begin{figure}[t!]
    \centering
    \includegraphics[width=0.5\linewidth]{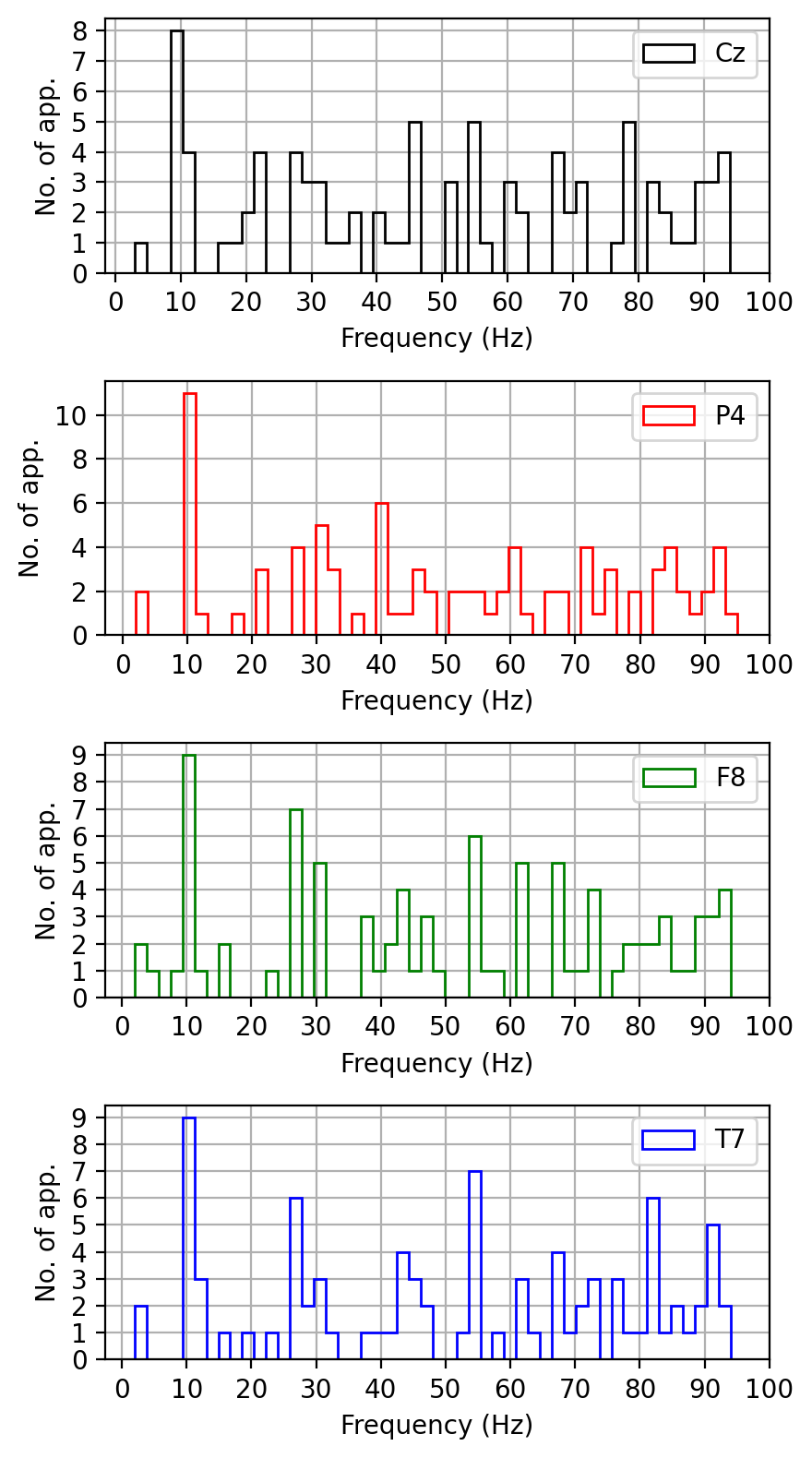}
    \caption{Histogram of the peak response frequency of the EFR of Cz, F8, P4, and T7 channels for 13 participants.}
    \label{fig:Hist_Tf_peak_freq}
\end{figure}

\autoref{fig:Hist_Tf_peak_freq} illustrates the distribution of dominating frequency for the transfer functions of Cz, P4, F8 and T7 channels. The histograms for Cz, P4, F8 and T7 show several peaks, indicating that multiple participants' experimental data contained frequencies within the bin. The Cz exhibits peaks at 9-10 Hz, 45-46 Hz, 54-55 Hz and 79-80 Hz, P4 at 10-11 Hz and 40-41 Hz, F8 at 10-11 Hz, 26-27 Hz, 30-31 Hz and 54-55 Hz, and T7 at 10-11 Hz, 26-27 Hz, 54-55 Hz, 81-82 Hz, and 91-92 Hz. The frequency ranges \emph{8-11 Hz}, \emph{53-56 Hz}, and \emph{78-81 Hz} have higher counts of peak occurrences across participants, with up to 10 instances in some bins. These commonly appearing frequencies produce strong neural responses across different brain regions and participants. The comprehensive view of the dominant frequencies and their distribution highlights the frequency-specific neural processes for the underlying cognitive or sensory functions. The histogram also shows additional inconsistent peak response frequencies for all four channels. These observations suggest the 8-11 Hz frequency band is the common resonance frequency of the Cz, P4, F8 and T7 channels for the EFR. This band focuses on auditory stimuli by suppressing background noise and irrelevant inputs \cite{foxe2011role, Klimesch2012}. Gamma oscillations in the lower range (53--56 Hz) are involved in coherent perception by binding auditory features, such as phonemes and syllables. This frequency range is particularly relevant for processing rapid temporal changes in auditory stimuli and correlates with attributes at the phonemic scale, such as formant transitions \cite{poeppel2003analysis}. The high gamma responses are related to the representation of the neural object of the auditory stimuli \cite{palva2002distinct}. Besides this, this gamma activity indicates selective attention to a sensory stimulus \cite{ray2008high} and memory retrieval process \cite{herrmann2004cognitive}.
\subsection*{Frequently appeared dominant spatial-spectral cross-correlation}
\begin{figure}[t!]
    \centering
        \includegraphics[width=0.6\linewidth]{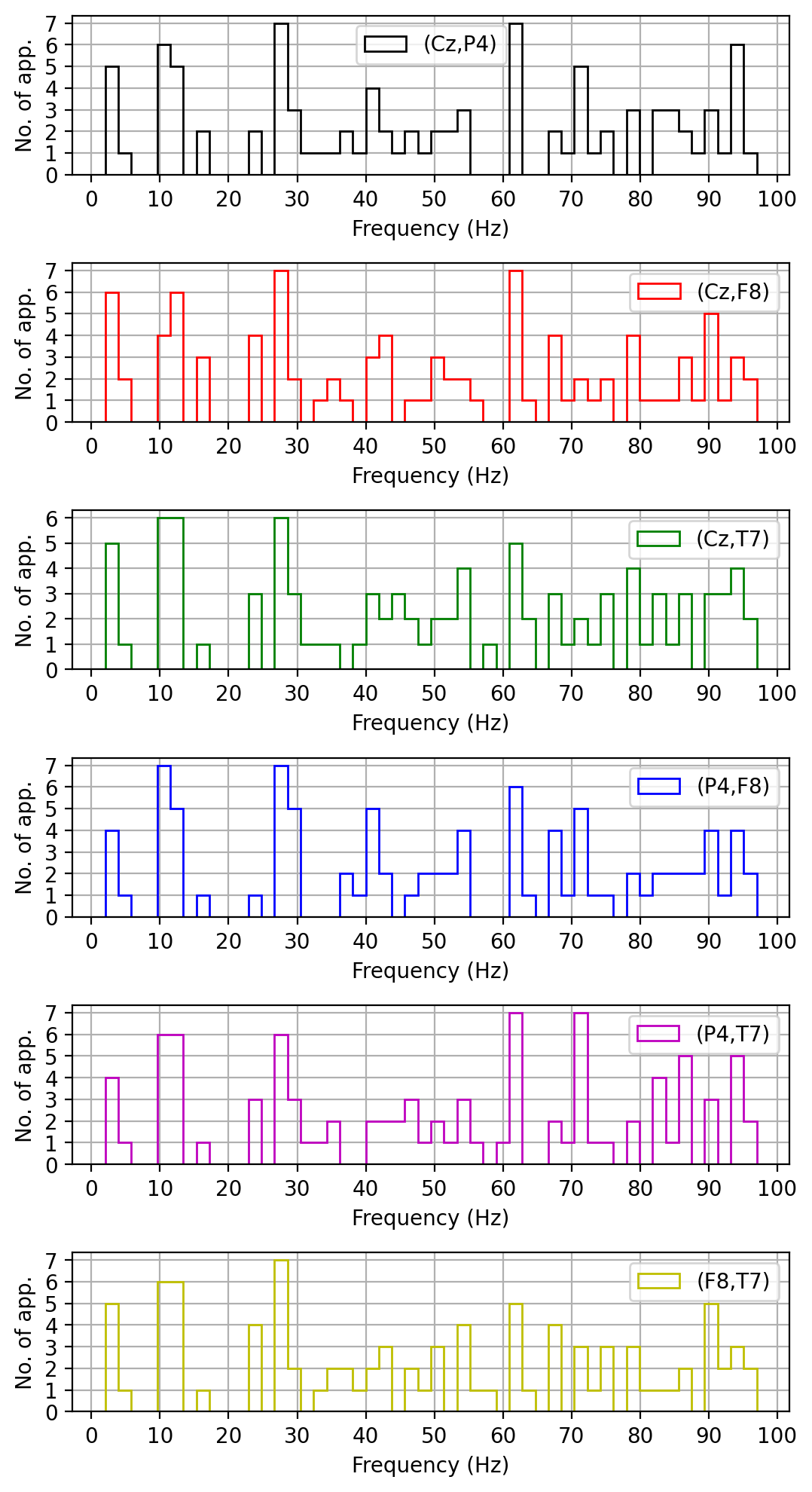}
        \caption{The appearance of dominating frequencies of the CSD functions for different for the 13 participants.}
    \label{fig:TfCr_hist}
\end{figure}

The histogram shown in \autoref{fig:TfCr_hist} represents the frequently appearing dominating frequencies of the CSD for the different channel pairs. The frequency-dependent cross-correlation indicates the synchronisation of dominating frequency responses across Cz, P4, F8, and T7 channels. A whole-brain intracranial EEG study demonstrates that auditory perception relies on distributed, frequency-specific coherent neural oscillations \cite{te2024speech}. 

The peaks of the (Cz, P4) pair are most frequent at around 9-11 Hz, 27-29 Hz, 62-64 Hz, and 93-95 Hz. The synchronisation in the alpha (9-11 Hz) band might reflect the coordination of attentional resources between central and parietal regions during auditory perception \cite{obleser2012alpha}. The strong neural connectivity or shared neural activity at the beta band (27-29 Hz) indicates involvement of sensorimotor areas in processing speech's rhythm and temporal structure \citep{bastos2012canonical}. Gamma responses (62-64 Hz and 93-95 Hz) are related to higher cognitive functions, including attention, memory, and perception integration. The CSD peaks in the gamma range suggest the joint involvement of these regions in integrating speech elements with higher cognitive functions \cite{fries2009neuronal}, \citet{canolty2006high}.


In the (Cz, F8) pair, cross-correlation peaks appear most frequently at around 2-4 Hz, 12-14 Hz, 27-29 Hz, and 62-64 Hz. The frequently appearing delta (2-4 Hz) CSD responses may reveal their coordinated role in speech envelope following \cite{brohl2021delta}. The synchronous activity of 12-14 Hz (the lower beta band) may correspond to the prediction and anticipation of auditory stimuli \cite{biau2018lower}. The spatial coherence found around 27-29 Hz likely reflects the auditory sensory processing and feature extraction \cite{betti2021spontaneous}. Lastly, the 62-64 Hz peak is associated with higher-level cognitive functions such as attention and working memory \cite{fries2001gamma}. 
 
The cross-correlation for the (Cz, T7) pair has peaks at 10–13 Hz, 27–29 Hz, and 62–64 Hz. Like other channel pairs, the spatial synchronisation at 10-13 Hz, alpha band, is related to the suppression of irrelevant auditory input, allowing the brain to focus on salient features \cite{foxe2011role}. The peak at 27-29 Hz has been linked to alertness, attention, and emotional arousal. The gamma-band peaks (62–64 Hz) reveal the integration of phonological and syntactic features of speech \cite{giraud2012cortical}. 
 
For the (P4, F8) pair, the most frequent cross-correlation peaks are located at approximately 10–12 Hz, 27–29 Hz, and 62–64 Hz. The spatial synchronisation for the (P4, F8) pair is similar to the (Cz, T7) pair. The (P4, T7) pair exhibits frequent cross-correlation peaks at 10–13 Hz, 27–29 Hz, 62–64 Hz, and 71–73 Hz. This pair has a similar frequency-dependent spatial synchronisation pattern like the (Cz, T7) pair, with an additional synchronisation at 71–73 Hz. Notably, the additional peak in the gamma range (71–73 Hz) may reflect heightened neural synchronisation and increased cognitive load, aligning with studies on the role of gamma oscillations in demanding auditory processing tasks \cite{canolty2006high, zion2012cocktail}. 
 
Finally, the (F8, T7) pair demonstrates peak cross-correlations at 10–13 Hz, 27–29 Hz, and 62-64 Hz. Synchronisation in the \emph{alpha band (10-13 Hz)}, \emph{beta band (27- 29 Hz)} and \emph{gamma band (62-62 Hz)} have also found in (Cz, P4), (Cz, F8), (Cz, T7), (P4, F8), (P4, T7), and (F8, T7) pair which indicates this synchronisation is common. 

\subsection*{Conclusion}
This study explored the frequency-dependent neural responses excited by the auditory stimulus envelope using EEG signals recorded from multiple channels (Cz, P4, F8, and T7). The analysis revealed three significant frequency responses and three dominant spatial neural coherence across different brain regions, providing insights into the functional connectivity and neural processing of auditory stimuli. The time-averaged frequency responses highlighted that dominant peaks frequently appeared in the \emph{alpha (8-11 Hz)}, \emph{lower gamma (53-56 Hz)}, and \emph{higher gamma (78-81 Hz)} bands. These frequency responses may play dominant roles in speech perception by engaging and maintaining attention, suppressing irrelevant inputs, binding acoustic features, and memory processing. Cross-correlation of the transfer functions between channel pairs reveals the frequency-dependent neural synchronisation patterns. The frequently appearing neural coherences for all the channel pairs are \emph{10–13 Hz}, \emph{27–29 Hz}, and \emph{62–64 Hz}, indicating functional connectivity across cortical regions engaged in speech perception. The frequency-specific functional connectivity reflects attentional modulation, irrelevant input suppression, integration of acoustic features, and neural synchronisation in auditory perception. These frequency-specific findings reveal the key frequency bands and neural mechanisms of auditory perception.


\section*{Methods}


We aim to analyse speech envelopes following a response (EFR)-generating system, which necessitates estimating the envelope function of the speech as a first step. We estimate the envelope of the speech using the following approach:
\begin{enumerate}
    \item Filter the speech signal: $s(t)$ is filtered with a low-pass FIR filter having a cut-off frequency of 1000 Hz, transition width of 30 Hz, pass-band ripple of 0.00175 dB, stop-band attenuation of 60 dB
    \item Compute the analytic signal: $z(t) = s(t) + jH[s(t)]$; where $H$ is the Hilbert transform
    \item Calculate the instantaneous amplitude: $A(t) = |z(t)|$
    \item The envelope function is further filtered with a low-pass FIR filter, having a transition width of 20 Hz, pass-band ripple of 0.00175 dB and stop-band attenuation of 60 dB. Following \citet{aiken2008human}, the cut-off frequency is chosen as 100 Hz because it encompasses the primary frequency components associated with speech rhythms, such as syllables and prosody.
\end{enumerate}

The EFR-generation system is time-varying, and so we estimate the system transfer function for a short time using STFT, for which the system is considered time-invariant. The first step is \textit{segmentation}, where we divide the signals $a(n)$ and $e(n)$ into overlapping frames using a window function $w(n)$ with hop size $L$. In the second step, we \textit{apply STFT} to each frame of $a(n)$ and $e(n)$ to obtain their time-frequency representations $A(m,k)$ and $E(m,k)$. Finally, to \textit{estimate transfer function} $H(m,k)$, we calculate the ratio of $E(m,k)$ to $A(m,k)$ at each time-frequency point:
\begin{equation}
\begin{split} \label{eqn:TF}
    H(m,k) &= \frac{E(m,k)}{A(m,k)} \\
            &= \frac{\sum_{n=0}^{N-1} e(n+mL) w(n) e^{-j2\pi kn/N}}{\sum_{n=0}^{N-1} a(n+mL) w(n) e^{-j2\pi kn/N}}
\end{split}
\end{equation} 
where $H(m,k)$ is a discrete transfer function at time frame $m$ and frequency bin $k$, $E(m,k)$ is STFT of the EFR $e(n)$, $A(m,k)$ is STFT of the speech envelope $a(n)$, $N$ is STFT window length, $L$ is hop size between adjacent STFT frames and $w(n)$ is STFT window function.

The transfer function encapsulates how a specific EEG channel responds to a speech envelope, $A(n)$. It characterises the channel's filtering properties regarding amplitude and phase modifications across frequencies. The magnitude of the transfer function, $|H(m,k)|$, gives the response at frequency $f_k$ and time $t_m$. A higher magnitude indicates that the signal's energy at that frequency and time is more effectively transmitted through the system, while a lower magnitude indicates attenuation.

Propagation delay is associated with the phase difference between EFR and the speech envelope. The phase of the input and output signals is extracted and unwrapped:
\begin{align}
\phi_{\text{A}}(m, k) = \arg(A(m,k) \\
\phi_{\text{E}}(m, k) = \arg(E(m,k)
\end{align}

The phase difference between the output and input is calculated and constrained within the range \([0, 2\pi)\):
\begin{equation}
    \Delta \phi(m, k) = \phi_{\text{E}}(m, k) - \phi_{\text{A}}(m, k)
\end{equation}

The phase delay is computed as the time-averaged phase difference for each frequency bin:
\begin{equation}
\overline{\Delta \phi( k)} = \frac{1}{M} \sum_{m} \Delta \phi(m, k)
\end{equation}
where $M$ is the number of time indices.

Changes in $|H(m,k)|$ over time can indicate dynamic changes in the system that could affect the synchronisation or desynchronisation of neural responses in an auditory system.

The cross-correlation of transfer functions assesses the degree of similarity or synchrony between the dynamic responses of two EEG channels. The mathematical relation of the cross-spectral density (CSD) of two EEG channels, $H_x(m,k)$ and $H_y(m,k)$, is
\begin{equation}
    \begin{split}\label{eq: TfCr}
        R_{xy}(\tau, k) &= \sum_{m=-\infty}^{\infty} H_x(m,k) H_y^*(m+\tau, k) \\
        &= \frac{1}{M} \sum_{m=0}^{M-1} H_x(m,k) H_y^*(m+\tau, k)  
    \end{split}
\end{equation}
where $R_{xy}(\tau,k)$ is CSD at time lag $\tau$ and frequency bin $k$, $M$ is the number of time frames considered for cross-correlation and $^*$ refers to the complex conjugate operator.

We can write the polar form of the transfer functions as:
\begin{equation}
    \begin{split}\label{eq: Tfplr}
        &H_x(m,k)=|H_x(m,k)|e^{j\phi_x(m,k)}\\
        &H_y^*(m, k)=|H_y(m, k)|e^{-j\phi_y(m,k)}
    \end{split}
\end{equation}

Subsequently, we can write the expression of CSD as:
\begin{equation}
    \begin{split}\label{eq: TfCrmod}
        R_{xy}(\tau, k) &= \frac{1}{M} \sum_{m=0}^{M-1} |H_x(m,k)|e^{j\phi_x(m,k)} |H_y(m+\tau, k)|e^{-j\phi_y(m+\tau,k)} \\
        &=\frac{1}{M} \sum_{m=0}^{M-1}|H_{xy,m,\tau}(k)|e^{\phi_{xy,m, \tau}(k)}
    \end{split}
\end{equation}
where $|H_{xy,m,\tau}(k)|=|H_x(m,k)||H_y(m+\tau, k)|$ and $\phi_{xy,m, \tau}(k)=e^{j(\phi_x(m,k)-\phi_y(m+\tau,k))}$ are the multiplication of the magnitude responses and the phase difference. 

The consistency of the phase difference is the key factor for CSD. If the phase difference is consistent, $\phi_{xy,m, \tau}(k)=\phi_{xy, \tau}(k)$, then \autoref{eq: TfCrmod} becomes:
\begin{equation}\label{eq: TfCrphconsistent}
     R_{xy}(\tau, k) = e^{\phi_{xy, \tau}(k)}\frac{1}{M} \sum_{m=0}^{M-1}|H_{xy,m,\tau}(k)|
\end{equation}
where the summation of the magnitudes, $|H_{xy,m}(k)|$ contributes constructively to $R_{xy}(\tau, k)$. Conversely, if $\phi_{xy,m, \tau}(k)$ varies, the exponential term oscillates, leading to potential cancellation effects during summation, thus reducing $|R_{xy}(\tau, k)|$.

We can write the zero-lag CSD as:
\begin{equation}\label{eq: TfCrzero}
     R_{xy}(k) = \frac{1}{M} \sum_{m=0}^{M-1}|H_{xy,m}(k)|e^{j\phi_{xy,m}(k)}
\end{equation}

Then, the cross-spectral coefficient between the EFRs collected at two different EEG electrodes can be represented as:
\begin{equation}
\begin{split}
C_{xy}(k) &= \frac{\sqrt{|R_{xy}(0, k)|^2}}{\sqrt{R_{xx}(0, k) R_{yy}(0, k)}} \\
&= \frac{\sqrt{|\sum_{m=0}^{M-1} H_x(m,k) H_y^*(m, k)|^2}}{\sqrt{\sum_{m=0}^{M-1} |H_x(m,k)|^2 \sum_{m=0}^{M-1} |H_y(m,k)|^2}}
\end{split}
\end{equation}
where $C_{xy}(k)$ is the coherence between the transfer functions at frequency bin $k$, $R_{xy}(0,k)$ is the cross-correlation between the transfer functions at zero time lag and frequency bin $k$, $R_{xx}(0,k)$ is the auto-correlation of the transfer function of channel $x$ at zero time lag and frequency bin $k$ and $R_{yy}(0,k)$ is the auto-correlation of the transfer function of channel $y$ at zero time lag and frequency bin $k$.

\subsection*{Sound and EEG Data}\label{sec:database}
We sourced data from a publicly available dataset \cite{alzahab2022auditory} available on PhysioNet \cite{goldberger2000physiobank}, which includes resting state electroencephalographic (EEG) signals recorded during auditory stimulation. The EEG signals were captured using the OpenBCI GUI (v5.0.3) and an OpenBCI Ganglion Board, sampling four channels (T7, F8, Cz and P4) at 200 Hz. Four gold-cup electrodes with Ten20 Conductive Paste were used for data acquisition, with reference and ground electrodes placed on the left and right ears, respectively. Before recording, skin impedance was measured to ensure proper electrode connectivity. For our analysis, we utilised the recordings of subjects exposed to native songs, with stimuli presented through in-ear headphones. We selected EEG data from 13 listeners (S01, S02, S03, S04, S06, S08, S09, S10, S11, S12, S13, S14, and S15). Their demographic information is provided in the "Subject.csv" file available in \cite{alzahab2022auditory}. The stimuli consisted of Arabic songs for Arabic-speaking listeners and Italian songs for Italian-speaking listeners. We extracted two-minute segments of EEG data from the raw recordings, aligning them with the auditory stimuli. The stimuli were sourced from YouTube, trimmed to two minutes, and converted from stereo to mono-channel for analysis.

\newpage


\section*{Resource availability}


\subsection*{Lead contact}


Requests for further information and resources should be directed to and will be fulfilled by the lead contact, Md. Mahbub Hasan (mahbub01@eee.kuet.ac.bd).

\subsection*{Materials availability}


This study did not generate new materials.

\subsection*{Data and code availability}


\begin{itemize}
    \item The primary data used in this paper are accessed from a publicly available dataset \cite{alzahab2022auditory} hosted on PhysioNet \cite{goldberger2000physiobank} as of the date of publication.
    \item All code has been deposited at a GitHub repository (\url{https://github.com/hasan-rakibul/speech-efr-eeg}) and is publicly available.
    \item Any additional information required to reanalyse the data reported in this paper is available from the lead contact upon request.
\end{itemize}

\section*{Acknowledgments}


This work was funded by CASR, Khulna University of Engineering \& Technology, Khulna 9203, Bangladesh.

\section*{Author contributions}


Conceptualization, M.M.H.; methodology, M.M.H.; investigation, M.M.H.; writing-–original draft, M.M.H. and M.R.H.; writing-–review \& editing, M.R.H., M.M.H., M.Z.H. and T.G.; funding acquisition, M.M.H.; resources, M.M.H. and M.R.H.; supervision, M.M.H.

\section*{DECLARATION OF INTERESTS}


The authors declare no competing interests.




\section*{SUPPLEMENTAL INFORMATION INDEX}




\begin{description}
  \item Table S1: The frequencies of peak magnitude response of the transfer functions of Cz, P4, F8, and T7 channels. 
(\href{Table-S1.pdf}{PDF})
\end{description}

\bibliography{references}

\end{document}


\begin{table}[htbp]
\centering
\caption{Frequencies at which the maximum responses of the transfer function have been obtained for different subjects}
\label{APP_tab:Peak_tf_frequency}
\begin{tabularx}{\linewidth}{|c|X|X|X|X|}
\hline
 \textbf{Subject} & \textbf{Tf\_Cz} & \textbf{Tf\_P4} & \textbf{Tf\_F8} & \textbf{Tf\_T7} \\
\hline
         1 & [10.0, 22.0, 32.0, 46.0, 61.0, 76.0, 88.0]       & [11.0, 32.0, 47.0, 61.0, 76.0, 89.0]             & [5.0, 16.0, 27.0, 41.0, 55.0, 67.0, 80.0, 92.0]  & [16.0, 32.0, 47.0, 62.0, 76.0, 86.0]            \\
 \hline
       2 & [10.0, 27.0, 37.0, 52.0, 62.0, 72.0, 84.0]       & [10.0, 22.0, 32.0, 47.0, 61.0, 72.0, 84.0]       & [10.0, 27.0, 37.0, 47.0, 62.0, 76.0, 92.0]       & [10.0, 27.0, 37.0, 47.0, 64.0, 76.0, 92.0]      \\
  \hline
          3 & [3.0, 17.0, 30.0, 42.0, 54.0, 69.0, 82.0, 94.0]  & [2.0, 17.0, 30.0, 41.0, 55.0, 67.0, 82.0, 92.0]  & [2.0, 16.0, 30.0, 47.0, 57.0, 68.0, 80.0, 90.0]  & [12.0, 30.0, 43.0, 55.0, 67.0, 80.0, 90.0]      \\
  \hline
          4 & [11.0, 23.0, 35.0, 55.0, 72.0, 82.0, 94.0]       & [11.0, 30.0, 41.0, 55.0, 72.0, 90.0]             & [11.0, 30.0, 44.0, 55.0, 72.0, 90.0]             & [2.0, 12.0, 28.0, 43.0, 55.0, 72.0, 82.0, 92.0] \\
  \hline
          5 & [12.0, 30.0, 44.0, 54.0, 67.0, 82.0, 92.0]       & [2.0, 12.0, 30.0, 41.0, 54.0, 72.0, 82.0, 92.0]  & [2.0, 12.0, 30.0, 44.0, 54.0, 68.0, 82.0, 92.0]  & [2.0, 12.0, 30.0, 44.0, 55.0, 67.0, 82.0, 94.0] \\
  \hline
        6 & [10.0, 21.0, 32.0, 45.0, 61.0, 79.0, 92.0]       & [11.0, 21.0, 32.0, 42.0, 61.0, 71.0, 84.0]       & [11.0, 27.0, 43.0, 54.0, 70.0, 84.0]             & [11.0, 27.0, 41.0, 54.0, 71.0, 82.0, 93.0]      \\
  \hline
         7 & [10.0, 23.0, 33.0, 55.0, 67.0, 79.0, 90.0]       & [10.0, 30.0, 44.0, 58.0, 68.0, 79.0, 90.0]       & [10.0, 30.0, 42.0, 55.0, 68.0, 82.0, 94.0]       & [10.0, 20.0, 30.0, 55.0, 68.0, 79.0, 90.0]      \\
  \hline
          8 & [10.0, 27.0, 45.0, 62.0, 79.0, 93.0]             & [10.0, 27.0, 45.0, 61.0, 76.0, 86.0]             & [10.0, 27.0, 45.0, 61.0, 71.0, 88.0]             & [10.0, 27.0, 45.0, 61.0, 71.0, 88.0]            \\
  \hline
         9 & [9.0, 19.0, 30.0, 41.0, 55.0, 72.0, 90.0]        & [10.0, 30.0, 41.0, 54.0, 67.0, 82.0, 92.0]       & [9.0, 30.0, 54.0, 67.0, 79.0, 90.0]              & [10.0, 28.0, 43.0, 54.0, 67.0, 82.0, 92.0]      \\
  \hline
        10 & [11.0, 21.0, 32.0, 52.0, 70.0, 84.0, 94.0]       & [11.0, 27.0, 40.0, 52.0, 62.0, 76.0, 86.0]       & [11.0, 27.0, 37.0, 48.0, 61.0, 72.0, 84.0, 94.0] & [11.0, 27.0, 52.0, 76.0, 86.0]                  \\
  \hline
        11 & [10.0, 27.0, 40.0, 52.0, 68.0, 86.0]             & [10.0, 27.0, 41.0, 52.0, 68.0, 84.0]             & [11.0, 27.0, 37.0, 47.0, 61.0, 72.0, 84.0, 94.0] & [10.0, 27.0, 45.0, 61.0, 72.0, 82.0, 92.0]      \\
 \hline
        12 & [11.0, 23.0, 36.0, 46.0, 56.0, 67.0, 79.0, 90.0] & [11.0, 22.0, 36.0, 46.0, 56.0, 73.0, 84.0, 95.0] & [11.0, 23.0, 40.0, 59.0, 73.0, 86.0]             & [11.0, 23.0, 40.0, 55.0, 73.0, 84.0]            \\
 \hline
       13 & [10.0, 27.0, 45.0, 61.0, 79.0, 92.0]             & [10.0, 27.0, 45.0, 59.0, 79.0, 92.0]             & [10.0, 27.0, 43.0, 61.0, 79.0, 94.0]             & [10.0, 27.0, 45.0, 59.0, 70.0, 82.0, 92.0]      \\
\hline
\end{tabularx}
\end{table}